# PYROELECTRIC ORIGIN OF THE CARRIER DENSITY MODULATION AT GRAPHENE-FERROELECTRIC INTERFACE

Anna N. Morozovska[1,2], Maksym V. Strikha[*2]

[1]Institute of Physics, NAS of Ukraine, 46, pr. Nauky, 03028 Kyiv, Ukraine,

[2]V.Lashkariov Institute of Semiconductor Physics, NAS of Ukraine,
41, pr. Nauky, 03028 Kyiv, Ukraine,

## Abstract

Using continuous approximation we study the static and high-frequency heat dissipation in multi-layer graphene on a ferroelectric. We demonstrate that the Joule heating effect, caused by a high-frequency *ac* electric current in graphene, creates a pronounced temperature gradient in a ferroelectric substrate. The pyroelectric effect transforms the gradient into the spontaneous polarization gradient.. Therefore, the high-frequency depolarizing electric field occurs and penetrates in the multi-layer graphene. Free charges in graphene immediately screen the electric field and thus their density oscillates at high-frequency. Performed calculations had proved that the pyroelectric effect can modify essentially the free carrier density at the graphene-ferroelectric interface and consequently the conductivity of multi-layer graphene channel. So, pyroelectric mechanism can be critical for understanding of the complex physical thermal and electrical processes taking place across and along graphene-ferroelectric interfaces at terahertz frequencies.

## I. INTRODUCTION

Electronic, electromechanical, thermal, optical and other physical properties of graphene are widely and actively studied (see e.g. works [1, 2, 3, 4, 5, 6, 7] and refs therein). The physical problem of heat transfer across monolayer and few layer graphene can be of principal importance for the graphene-on-substrate based devices, operating at GHz frequencies [8, 9], because the Joule heating can modify substantially the system characteristics, and the role of interfaces and substrate choice can be crucial.

Kim et al studied [10] the Joule heating effect on graphene electronic properties. A number of technologically important substrate materials such as $SiO_2$, SiC, hexagonal BN, and diamond were taken into consideration. The results illustrate that the choice of substrate has a major impact via the heat transfer and surface polar phonon scattering. Particularly, it was found that the poor thermal conductivity of $SiO_2$ leads to significant Joule heating and saturation velocity degradation in graphene (characterized by the so-called $1/\sqrt{n}$ decay with the carrier concentration $n$).

* maksym_strikha@hotmail.com

Recent experimental and theoretical studies consider intriguing physical properties of graphene placed on ferroelectric substrates, both organic ferroelectric relaxors and inorganic ceramic $PbZr_xTi_{1-x}O_3$ (PZT) [11, 12, 13, 14, 15, 16]. Interest to ferroelectric substrates is primary related with their high static dielectric permittivity, that is of order $10^2-10^3$ for PZT [17] and $10^3-10^5$ for ferroelectric relaxors [18], as well as with the prominent possibility to add next level of functionality by electric field and temperature control over the spontaneous polarization direction, value and domain structure properties in the vicinity of surface [19, 20, 21, 22]. Since all ferroelectric materials used in experiments [11-16] are wide-band gap semiconductors or even dielectrics [23], a strong depolarization electric field, caused by the spontaneous polarization abrupt disappearance at the ferroelectric surface [17-23], should be screened by the carriers localized at the graphene-ferroelectric interface, where the type of the carriers (positive or negative) is defined by the spontaneous polarization direction. Piezoelectric and pyroelectric effects inherent to all ferroelectrics allow strong and controllable modification of the spontaneous polarization properties by external elastic stress and temperature variation [17-23]. Consequently ferroelectric substrate can make graphene interface highly sensitive to external stimuli such as electric field (polar-active), strain (piezo-active) and temperature (pyro-active).

Thus the system graphene-on-ferroelectric becomes “smart” and can acquire unique advantages in comparison with the graphene deposited on $SiO_2$ or high-k dielectrics (see rev. [24] and refs therein). In particular, one can obtain high ($\sim 10^{12}$ cm$^{-2}$) carrier concentration in the doped graphene-on-ferroelectric structures for moderate (of the order of 1 V) gate voltages. Existence of a hysteresis (or anti-hysteresis) in the dependence of graphene channel electrical resistance on the gate voltage can facilitate creation of bistable systems with unique physical properties. Graphene-on-ferroelectric evidently demonstrates such (anti)hysteresis behavior [11-15]. Therefore ferroelectric substrates enable development of robust elements of non-volatile memory of a new generation [11-15]. These elements operate for more than $10^5$ switching cycles and store information for more than $10^3$ s. Such systems can be characterized by ultrafast switching rates (~ 10–100 fs, see [11-15, 25, 26]). Theoretical analysis also demonstrated that the structures graphene-on-PZT would result in developing efficient and fast small-sized modulators of mid-IR and near-IR radiations suitable for different optoelectronic applications [24-26, 27].

Despite the abovementioned advances, fundamental mechanisms of the ferroelectric interface influence on the graphene physical properties are not enough studied until now [24-27]. Only recently a quantitative model is proposed to explain the anti-hysteresis behaviour of graphene-on-ferroelectric substrate resistance on the gate voltage sweep [28]. The model takes into consideration a capture of electrons from graphene sheet by the states, connected with graphene-ferroelectric interface. However, high-frequency heat dissipation in graphene (caused by a Joule effect from a high-frequency ac electric current along graphene sheets) can create a pronounced temperature gradient in a ferroelectric substrate. The pyroelectric effect transforms the temperature gradient in a ferroelectric into the spontaneous polarization gradient. For the spontaneous polarization component perpendicular to the graphene-ferroelectric interface, the high-frequency depolarizing electric field occurs and penetrates in the multi-layer graphene. Free charges in graphene immediately screen the electric field and thus their density oscillates at high-frequency. The physical model seems relatively simple, but to the best of our knowledge, the pyroelectric mechanism of the carrier density modulation at graphene-ferroelectric interface was neither discussed previously nor studied theoretically or experimentally. The absence of the theory motivates us to perform calculations of the heat dissipation in multi-layer graphene-on-ferroelectric substrate and the carrier density modulation at graphene-ferroelectric interface caused by pyroelectric effect.

Original part of the paper is organized as follows: in Sec.II we develop the model for Joule heating of the few-layer graphene in continuous approximation and analyze its applicability limits, in Sec.III and IV we study the static heating and high-frequency temperature modulation across graphene layers originated under the heating by high-frequency ac field. Pyroelectric mechanism of the carrier density modulation in graphene is considered in details in Sec.V. Possible outcomes are discussed in Sec.VI.

## II. JOULE HEATING OF THE MULTILAYER GRAPHENE-ON-SUBSTRATE

### II.1. Continuous model and basis equations

Model system is presented in **Fig.1**. Graphene heating is caused by the electric current, induced by *ac* electric field $E(t)$ applied along *x* axis. The static gate voltage field, that dopes *N*-layer graphene with electrons and holes, is applied along z axis.

Joule heating of graphene occurs due to the ac electric current, in turn caused by the field $E(t)=E_0\sin(\omega_0 t)$, that creates Joule heat sources with density

$$q(t)=\sigma\left|E^2(t)\right|=\frac{\sigma E_0^2}{2}\left(1-\cos(2\omega_0 t)\right) \tag{1}$$

Where σ is bulk conductivity of multi-layer graphene. Below symbol "~" over a letter stands for its frequency Fourier image. In particular the heat sources

$$\tilde{q}(\omega)=\frac{\sigma E_0^2}{2}\left(\delta(\omega)-\frac{\delta(\omega-2\omega_0)}{2}-\frac{\delta(\omega+2\omega_0)}{2}\right), \tag{1b}$$

where $\delta(\omega)$ is a Dirac-delta function.

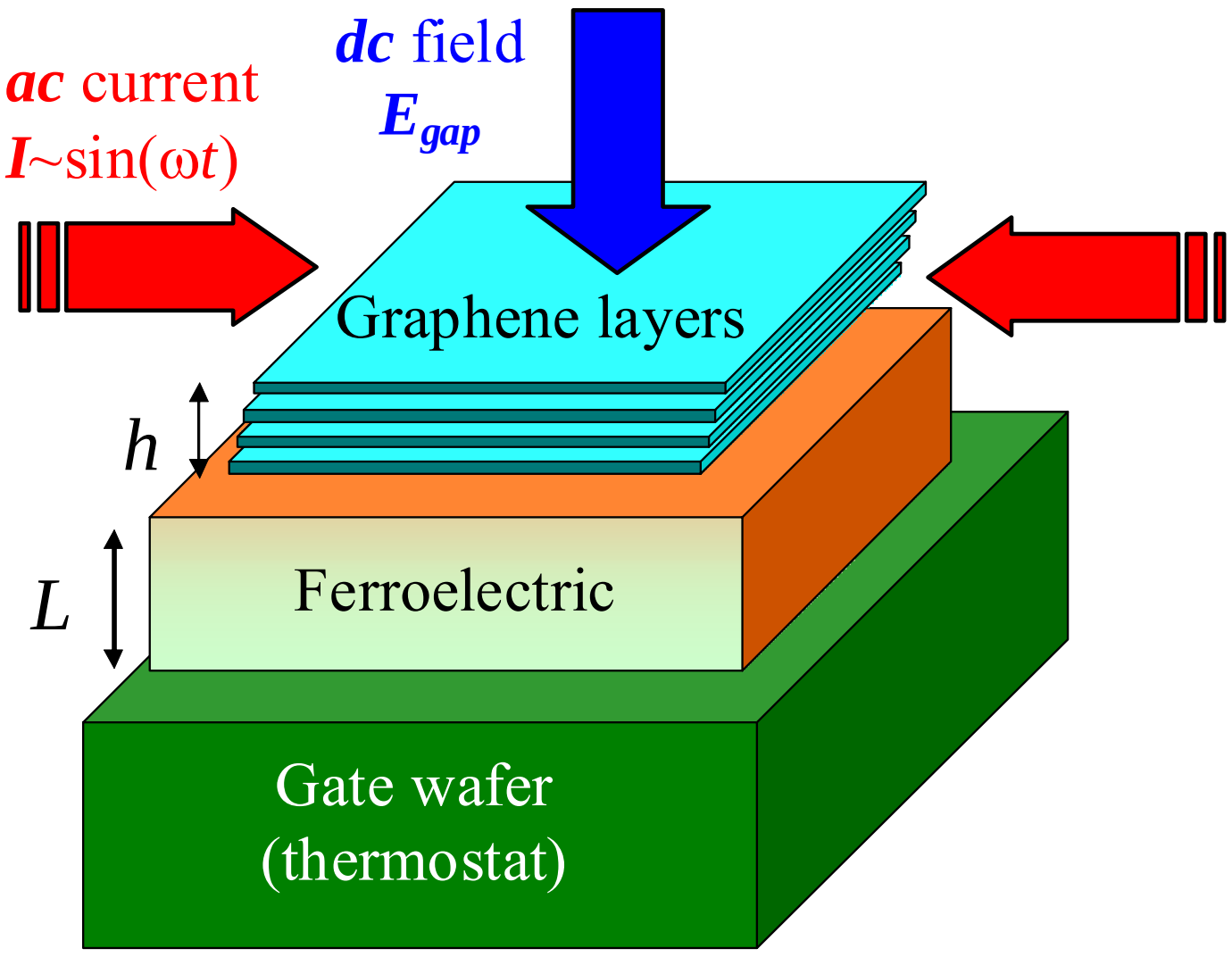

**Figure 1. Geometry of the problem**. *N*-layer graphene with thickness *h* on ferroelectric substrate of thickness L clamped to a gate wafer (thermostat).

Within a continuous theory (the limits of its application to *N*-layer graphene are discussed below) the hyperbolic equation for heat conductivity describing the variation of the temperature *T* distribution can be written as [29]:

$$\omega^2\tilde{T}_G - \frac{i\omega}{\tau}\tilde{T} + \frac{D_G}{\tau}\frac{d^2\tilde{T}_G}{d^2z} = -\frac{\tilde{q}}{c_V\tau}, \qquad 0 \le z \le h, \quad \text{(graphene)} \tag{2a}$$

$$-i\omega\tilde{T}_S + D_S\frac{d^2\tilde{T}_S}{d^2z} = 0, \qquad z > h \quad \text{(substrate)} \tag{2b}$$

where $c_V$ is a volumetric heat capacity.

The boundary conditions for Eqs. (2) are the thermal flux absence at the graphene-vacuum interface $z=0$; the frequency-dependent temperature jump $\Delta T$ and the thermal flux continuity on the graphene-substrate interface $z=h$ [30]; and unperturbed constant temperature field far from the interface in a gate wafer. Effect of nonzero $\Delta T$ can originate from the ultra-thin thermal gap at graphene-ferroelectric interface, leading to the disarrangement in phonon modes, and causing the temperature jump $\Delta T = R_H Q$, where $R_H$ is the Kapitza resistance [10, 31], $Q(z) = -\lambda_G\, dT_G/dz$ is the thermal flux. Thus the boundary conditions are:

$$\frac{d}{dz}\tilde{T}_G(0) = 0, \tag{3a}$$

$$\tilde{T}_G(h) - \tilde{T}_S(h) = -R_H\lambda_G\frac{d}{dz}\tilde{T}_G(h), \tag{3b}$$

$$\lambda_G \frac{d}{dz}\widetilde{T}_G(h) = \lambda_S \frac{d}{dz}\widetilde{T}_S(h), \qquad (3c)$$

$$\widetilde{T}_S(L+h) \to T_0. \qquad (3d)$$

Here $\lambda_G$ and $\lambda_S$ are the heat conductivity coefficients of graphene and substrate correspondingly, $T_0$ is the fixed temperature of a massive heat-conductive gate wafer. The parameters of heat transfer for the interfaces of graphene with different media were studied both experimentally and theoretically (see e.g. [10, 32, 33]). The contact with ferroelectric substrate, which is a very good heat conductor [34, 35], corresponds the case $\lambda_G << \lambda_S$. Estimations of $R_H$ will be made in the section II.3. Ideal heat contact, $\widetilde{T}_G(h) = \widetilde{T}_S(h)$, corresponds to the case $R_H = 0$. The high-frequency heat penetration depth is about several temperature wavelengths $\lambda_T \sim \omega_0^{-1/2}$ at high frequency $2\omega_0$. For the ferroelectric substrate with thickness $L$ ~100 nm the value of $\lambda_T$ is much smaller than $L$.

As it has been demonstrated experimentally, heat transfer across $N$-layer graphene. (later on we suppose $N >> 1$) is caused by phonons (see [10]). Therefore $\tau \sim l_p / v_s$ is a characteristic relaxation time of the process, where $v_s$ is a sound velocity across $N$-layer graphene, $l_p$ is a phonon free path length, $D_G = K/(c_V)$ is a thermal diffusion coefficient, $K$ is a thermal conductivity constant. Note, that in Eq.(1a) we neglect the heat transfer to metallic contacts, where the *ac* field is applied. This can be done for the reasons described by Freitag et al [36], who had demonstrated experimentally, that metallic contacts act as heat sinks, but not in a dominant fashion; while the heat-flow from the graphene to the gate oxide underneath is dominant.

Koh et al [32] had demonstrated that the heat transfer across the $N$-layer graphene occurs in ballistic regime, with phonons scattering on the graphene-vacuum and graphene-substrate interfaces. In the ballistic case Eq.(2a) can be rewritten as $\tau \frac{\partial^2 T_G}{\partial^2 t} = D_G \nabla^2 T_G + q\left(E^2(z,t)\right)$.

### II.2. Analytical solution

The solution of Eq. (2) with boundary conditions (3) was obtained using the integral transformations method as listed in **Appendix A, Suppl. Mat**. Below we are not interested in the transient process. In this case the stationary solution is the sum of a static component (subscript "0") and high-frequency temperature variation (subscript "ω") in graphene and in ferroelectric substrate:

$$T_G(z,t) = T_0 + \frac{\sigma E_0^2 h^2}{4 D_G c_V}\theta_0(z) + \frac{\left(\sigma E_0^2 \tau / 4 c_V\right)}{4\omega_0^2\tau^2 - 2i\omega_0\tau}\left(1 - \theta_\omega \cosh(k_G z)\right)e^{2i\omega_0 t} + c.c. \qquad (4)$$

$$T_S(z,t) = T_0 + \lambda\left(\frac{\sigma E_0^2 h(L+h)}{2 D_G c_V}\vartheta_0(z) + \frac{\left(\sigma E_0^2 \tau / 4 c_V\right)}{\left(4\omega_0^2\tau^2 - 2i\omega_0\tau\right)}\frac{k_G}{k_S}\theta_\omega \sinh(k_G h) e^{k_S(h-z)+2i\omega_0 t} + c.c.\right) \qquad (5)$$

Here $\lambda = \lambda_G / \lambda_S$ is the ratio of heat conductivities. In Eq.(4) $0 \le z \le h$ and $h \le z \le L$ in Eq.(5). Dimensionless functions $\vartheta_0(z)$, $\theta_0(z)$ and $\theta(\omega_0)$ are

$$\vartheta_0(z) = 1 - \frac{z}{L+h}, \quad \theta_0(z) = 1 - \frac{z^2}{h^2} + 2\lambda \frac{L}{h} + 2R_H \frac{\lambda_G}{h} \tag{6a}$$

$$\theta_\omega(\omega_0) = \frac{k_S}{k_S \cosh(k_G h) + \lambda k_G (1 + R_H k_S \lambda_S) \sinh(k_G h)} \tag{6b}$$

with the complex parameters

$$k_S = \frac{1+i}{2}\sqrt{\frac{2\omega_0}{D_S}} = \frac{1+i}{2l_D}\sqrt{2\omega_0\tau \frac{D_G}{D_S}}, \qquad k_G = \sqrt{\frac{2i\omega_0 - 4\omega_0^2\tau}{D_G}} \equiv \frac{\sqrt{2i\omega_0\tau - 4\omega_0^2\tau^2}}{l_D}. \tag{7}$$

Parameters $k_S$ and $k_G$ depend on dimensionless frequency $2\omega_0\tau$, ratio of diffusion coefficients $D_G/D_S$ and thermal diffusion length $l_D = \sqrt{D_G\tau}$.

In the ballistic limit we get $k_G(\omega) \approx \sqrt{-\omega^2\tau/D_G} \equiv i\omega\tau/l_D$, although generally the approximation $i\omega - \omega^2\tau^2 \to \omega^2\tau^2$ is valid for the high frequencies $\omega\tau >> 1$ only. No other formal simplifications occur in this case in Eq. (4). The pole in the equation (4) is possible in the hypothetic purely ballistic regime only, and the smallest addend from the diffusion term in Eq.(2a) helps us to avoid this non-physical pole.

Note, that the static heating depends on the product $R_H\lambda_G$, but not on the values $R_H$ and $\lambda_G$ separately. Physically the temperature jump $\Delta T$, which dominates in static heating, depends on the heat flux at the interface, which, in its term, depends on thermal conductivity of multi-layer graphene, where the Joule heating occurs. High-frequency temperature variation depends on the product $R_H\lambda_S$, but not on the values $R_H$ and $\lambda_S$ separately. Physically this variation is governed by $\Delta T$ jump and by heat dissipation into the substrate.

Mention, that the inequalities $D_G/D_S \ge 1$ and $\lambda_G/\lambda_S \ge 1$ correspond to the heat conductive graphene on less conductive ferroelectric substrate (equality corresponds to the same heat properties of graphene and substrate). The situation is improbable. The inequalities $D_G/D_S << 1$ and $\lambda_G/\lambda_S << 1$ correspond to the realistic situation of the less heat conductive graphene. Notice, that we treat the heat conductivity in multi-layer graphene across the layers, which is 2-3 orders of value smaller than in-plane heat conductivity of graphene [8], on more heat conductive substrate; which will be considered below.

### II.3. Estimations of the temperature variation amplitude and model applicability limit

The multi-layer graphene in-plain electric conductivity in Eq.(1) can be estimated as

$$\sigma(h) = e\mu n_{3D}(h), \tag{8a}$$

where the electron charge $e = 1.6\times10^{-19}$ C. Corresponding mobility $\mu \sim 1.4\times10^{1}$ m$^2$/Vs was observed experimentally for the 15-layer graphene on PZT (20/80) substrate in [13]. The bulk "effective" 3D-concentration $n_{3D}(h)$ is directly proportional to the 2D-concentration $n_{2D}$, determined for the gate-doped graphene by the gate voltage, and inversely proportional to the graphene thickness *h* (see e.g. [3, 27]):

$$n_{3D}(h) = \frac{n_{2D}}{h}. \tag{8b}$$

Due to ferroelectric high permittivity the value $n_{3D}(h) \sim 10^{18}\,\mathrm{m}^{-2}/h$ can be easily reached in graphene-on-ferroelectric for the moderate gate voltages [27]. It follows from Eqs.(8) that the factor $\sigma|E^2|/c_V$ included to the temperature variation (5) is inversely proportional to the *N*-layer graphene thickness *h* for the given gate voltage.

So, using the *ac* fields $E_0 \propto 5\times10^4$ V/m, heat capacity $c_V \approx 1.534\times10^6\ \mathrm{J/m^3K}$ [37], and taking into account that phonon relaxation time $\tau$ in graphene is of $10^{-11}$s order [38], the high-frequency temperature variation amplitudes in Eq.(5) can be estimated for mono-layer graphene (i.e. for the thickness $h_0 = 0.34$ nm [1]) as

$$\frac{\sigma E_0^2\tau}{4c_V} \equiv \frac{e\mu n_{2D}E_0^2\tau}{4hc_V} \propto \frac{1.6\times10^{-19}\mathrm{C}\left(1.4\times10^{19}\,/\mathrm{Vs}\right)\cdot\left(0.25\times10^{10}\mathrm{V^2/m^2}\right)\cdot10^{-11}\mathrm{s}}{0.34\cdot10^{-9}\mathrm{m}\cdot1.534\times10^{6}\ \mathrm{J/m^3K}} \approx 50\,\mathrm{K}. \tag{9a}$$

Equation (9a) gives about 5 K for 10-layer graphene ($h = Nh_0$).The thermal diffusion coefficient of graphene can be estimated as $D_G = K/c_V$ with coefficient $K \sim 2-20$ W/mK for the heat flux in graphite across the graphene layers [8] and $c_V \approx 1.534\times10^6\ \mathrm{J/m^3K}$ . This yields $D_G \approx (0.13-1.3)\cdot10^{-5}\ \mathrm{m}^2/s$ that is of the same order of values as $\left(10^{-6}-10^{-5}\right)\mathrm{m}^2/s$ for the single-wall carbon nanotubes [39].

The static heating in the 10-layer graphene for the ratio of heat conductivities $\lambda = \lambda_G/\lambda_S \sim 0.1$ can be estimated as:

$$\frac{\sigma E_0^2h^2}{4D_Gc_V} \equiv \frac{e\mu n_{2D}E_0^2\tau}{4c_V}\frac{h}{l_D^2} \propto 5\,\mathrm{K}, \tag{9b}$$

For the 100 nm-substrate:

$$\frac{\sigma E_0^2h(L+h)}{2D_Gc_V}\lambda \cong \frac{e\mu n_{2D}E_0^2\tau}{4c_V}\frac{L}{l_D^2}\lambda \propto 15\ \mathrm{K}. \tag{9c}$$

Here the diffusion length $l_D = \sqrt{D_G \tau}$ is introduced. Using relaxation time $\tau = 10^{-11}$ s we get the corresponding diffusion length $l_D \sim 3 - 10$ nm. That means that the continuum approximation [Eq.(2)] works quantitatively for $h \geq l_D/2$, i.e. for the number of graphene layers $N > 5$. Therefore numerical results for $N \geq 5$ presented in the next sections should be reliable.

Note, that the value of $E_0 = 5\times10^4$ V/m taken for estimations in Eq.(9) is consistent with approximations used above for the heat sources, because the transport in graphene is linear up to the fields of $\sim 2\cdot10^5$ V/m order (see e.g. [40]).

The amplitudes described by Eq.(9a, b, c) are inversely proportional to the multi-graphene graphene thickness due to the proportionality $n_{3D}(h) \sim 1/h$ in accordance with Eq.(8b). Since the amplitudes (9) can be of several Kelvins for $N$=10 at higher ac fields, the heating effect can not be regarded as negligible one.

Kapitza resistance for the interface graphene-quartz varies in the range from 5.6 $10^{-9}$ to 1.2 $10^{-8}$ Km$^2$/W [10]. First-principles calculations give 8.8 $10^{-9}$ Km$^2$/W [10]. It is reasonable to assume that $R_H$ value for the interface graphene-ferroelectric is about $R_H = \left(5\cdot10^{-9} - 10^{-8}\right)$ Km$^2$/W. Heat conductivity $\lambda_G \approx (1-10)$ W/(m·K) [10, 32], so the product $R_H \lambda_G = \left(5\cdot10^{-9} - 10^{-7}\right)$ m. Estimation for the dimensionless function $\theta_0(h) = 2\lambda\frac{L}{h} + 2R_H\frac{\lambda_G}{h}$ in Eq.(6a) for 10-layer graphene gives $30-80$. So that 10-layer graphene can be heated up to (5-100) K by the fields $E_0 = $(1-5)$\times10^4$ V/m correspondingly, since $\frac{\sigma E_0^2 h^2}{D_G c_V}\theta_0(h)$ reaches the temperatures for the chosen parameters.

Material parameters used in the estimations above and analytical calculations, which results are presented in the next sections, are summarized in the **Table SI, Suppl. Mat**.

## III. STATIC HEATING OF GRAPHENE-ON-SUBSTRATE

Analytical expressions for the static heating $\delta T = T - T_0$ of graphene and ferroelectric substrate are:

$$\delta T_G(z) = \frac{e\mu n_{2D} E_0^2}{4D_G c_V} h\left(1 - \frac{z^2}{h^2} + 2\lambda\frac{L}{h} + 2R_H\frac{\lambda_G}{h}\right), \tag{10a}$$

$$\delta T_S(z) = \lambda\frac{e\mu n_{2D} E_0^2}{2D_G c_V}(L + h - z), \tag{10b}$$

$$\left\langle \delta T_G(z)\right\rangle = \frac{1}{h}\int_0^h dz\,\delta T_G(z) \equiv \frac{e\mu n_{2D} E_0^2 h}{4D_G c_V}\left(\frac{2}{3} + 2\lambda\frac{L}{h} + 2R_H\frac{\lambda_G}{h}\right), \tag{10c}$$

$$\langle \delta T_S(z) \rangle = \frac{1}{L}\int_h^{L+h} dz\,\delta T_S(z) \equiv \lambda \frac{e\mu n_{2D}}{4D_G c_V} L . \qquad (10d)$$

In Eqs.(10) we took into account that the graphene conductivity is inversely proportional to its thickness, $\sigma(h) = e\mu n_{2D}/h$ . Static heating of graphene and ferroelectric substrate z-dependence is given by Eq.(10a,b). Temperature gradients in graphene and ferroelectric quadratically and linearly depend on the coordinate z correspondingly. The simple dependencies present a little interest to be thoroughly studied, meanwhile the average heating given by Eqs.(10c,d) does present a definite interest as reflecting the total effect.

One can see from Eqs.(10) that the static heating are proportional to the several combinations of parameters, namely the second power of electric field amplitude, $E_0^2$, the product $R_H\lambda_G$ and $\lambda L$. **Figure 2** demonstrates the dependence of the static heating on the parameters within their actual range, $R_H\lambda_G \propto \left(10^{-9} - 10^{-7}\right)$ m, $\lambda L \propto (1-50)$ nm, $E_0$ ~ (0.1 – 5)×10$^4$ V/m and $h \propto (1-20)h_0$ , where $h_0 = 0.34$ nm. *Z*-distribution of the temperature field is continuous at the graphene-substrate interface $z = h$ only for the case $R_H\lambda_G = 0$; for the realistic case $R_H\lambda_G \geq 1$ nm the pronounced jump is seen in the **Fig. 2a**. The average static heating of multi-layer graphene is obviously proportional to its thickness *h*. The static heating of ferroelectric is independent on the graphene thickness *h*, since graphene conductivity is inversely proportional to the thickness (see **Fig. 2b**). The average heating is proportional to the second power of electric field amplitude, $E_0^2$; the dependence of its amplitude on the field is shown in **Fig. 2c.** The average heating linearly increases with substrate thickness *L* increase (see **Fig. 2d**). In numbers the static heating of graphene layers and ferroelectric substrate appeared pronounced (2-20 K) and strongly dependent on the amplitude $E_0^2$ and thickness *L*.

The most interesting physical result presented in the section is the strong dependence of the graphene static heating on the value of interfacial heat resistance $R_H$ and the complete indifference of the ferroelectric heating on $R_H$ (compare Eq.(10a,c) with (10b,d)). The Kapitza resistance $R_H$ blocks all "extra" heat in graphene acting as effective heat gap. Physically this means the dominant role of the scattering of the graphene phonons on the interface (in agreement with the experimental data [32]). The optimal amount of heat (proportional to $\lambda L$) is determined by the heat flux continuity at the graphene-ferroelectric interface; it causes the linear temperature gradient in the ferroelectric film. Mention also, that the static heating of ferroelectric is independent on the graphene thickness.

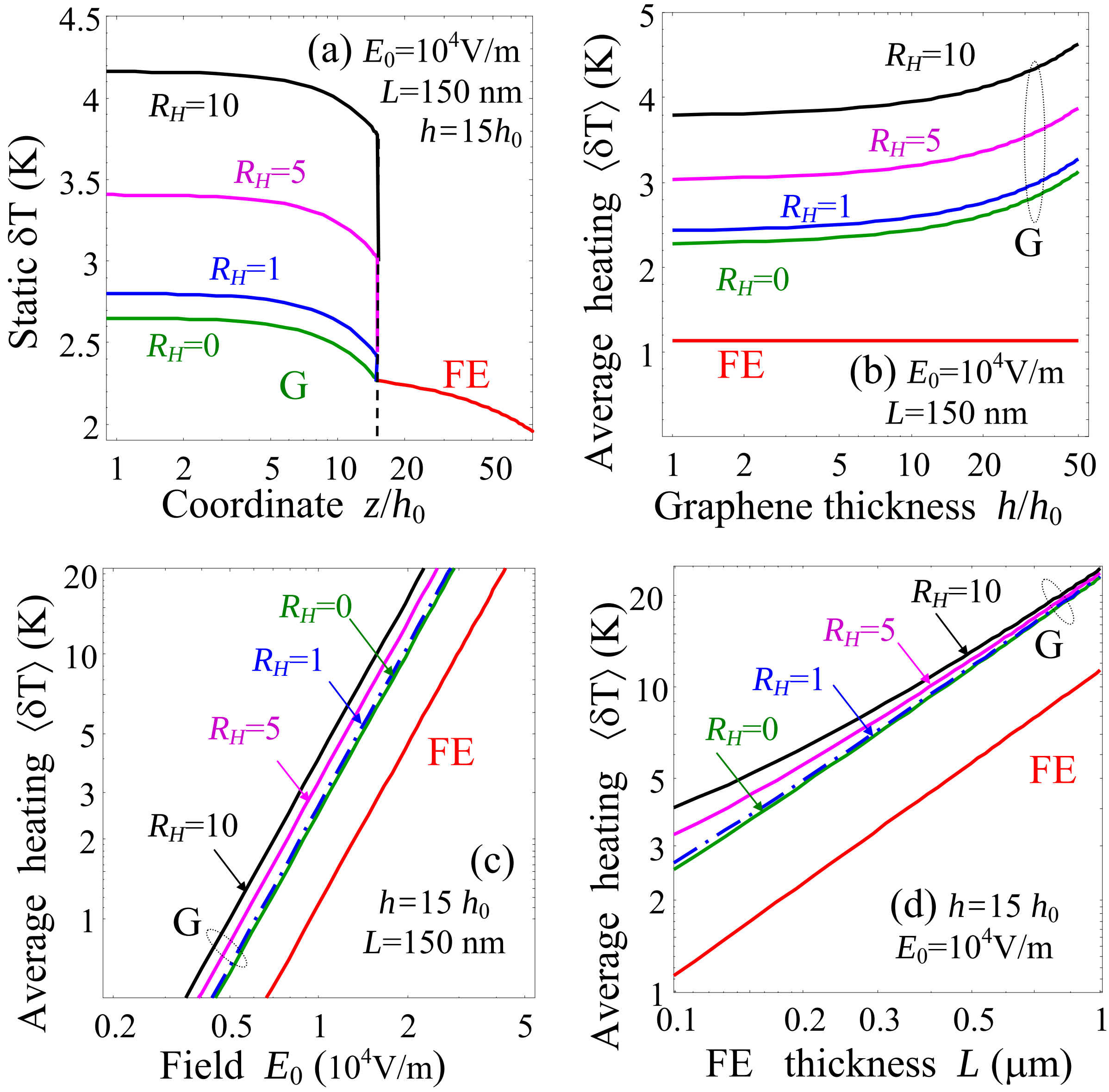

**Figure 2. Static heating of graphene-on-ferroelectric. (a)** Static heating $\delta T$ profile in multi-layer graphene (G) of thickness $h = 15h_0$ (5.1 nm) and ferroelectric substrate (FE-curves). The dependence of average heating $\langle \delta T \rangle$ on **(b)** graphene thickness $h/h_0$, **(c)** high-frequency electric field amplitude $E_0$ and **(d)** ferroelectric substrate thickness $L$. Curves are calculated for parameters $D_G/D_S$ =0.1, $\lambda_G/\lambda_S$ = 0.1, $l_D \approx 3$ nm, $L$ = 150 nm, $\lambda_G = 1$ W/(m·K) and $E_0$ = $10^4$ V/m (a, b, d). Interfacial resistance $R_H = (0,\ 1,\ 5,\ 10)\times 10^{-9}$ Km$^2$/W (as labeled near the curves). Other material parameters are listed in the **Table SI**.

## IV. HIGH-FREQUENCY MODULATION OF GRAPHENE TEMPERATURE

The high-frequency modulation of multi-layer graphene temperature appeared nontrivial and will be analyzed below. High-frequency temperature variations, calculated according to Eq. (5), are presented in **Figure 3** for several dimensionless frequencies $\omega_0\tau$, typical values of Kapitza resistance, $R_H = 10^{-8}$ Km$^2$/W and $R_H$ = 0 for comparison, heat conductivity $\lambda_G \approx$ 1 W/(m×K). The relaxation time $\tau \propto 10^{-11}$ s means that the dimensionless frequencies range $\omega\tau = 10$ - $10^{-1}$ correspond to the actual frequency from 1 THz to 10 GHz. Material parameters are listed in **Table SI.**

**Figures 3a** and **3b** demonstrate the dependence of the multi-layer graphene average temperature absolute value $\left|\langle\delta T_G(z)\rangle\right|$ and phase $\mathrm{Arg}\left(\langle\delta T_G(z)\rangle\right)$ on the thickness $h/h_0$ , i.e. on the number $N$ of graphene layers. Modulation profiles and average values decrease with the thickness increase and scale as $1/h$ for high thicknesses. At high frequencies the temperature modulation becomes small (~0.01 - 0.1 K) and weakly coordinate-dependent corresponding to a physically obvious limit of the frequency de-modulation. The modulation amplitude increases with the frequency decrease and reaches several Kelvins at frequencies $\omega\tau$ about or less than 1 (when the phonon relaxation is much slower than the ac field change) for multi-layer graphene thickness less than 20 layers. The modulation indeed reaches tens Kelvins for $N \leq 20$ and frequencies $\omega\tau$ equal or less than 0.1. Therefore modulation effect can not be regarded negligible in the frequency range. Dependences of temperature on coordinate $z$ for 15-layer graphene with two interfaces, vacuum and ferroelectric, correspondingly are shown in **Figures S1**, **Suppl. Mat.**

Ideal heat contact at graphene-ferroelectric interface corresponds to the case $R_H = 0$ . Corresponding high-frequency temperature modulation is presented in **Figs.3c,d** for the same dimensionless frequencies and other parameters as in **Fig. 3a,b**. Remarkable differences between the temperatures shown in the **Fig.3a,b** ($R_H = 0$) and in the **Fig.3c,d** ($R_H \neq 0$) are the following. The amplitude $\left|\langle\delta T_G(z)\rangle\right|$ is about an order of magnitude higher for the case $R_H = 10^{-8}$ Km$^2$/W than for the case $R_H = 0$ at moderate frequencies $\omega\tau \leq 1$. Dependence of the average value $\left|\langle\delta T_G(z)\rangle\right|$ on graphene layer thickness $h$ is monotonic and scales as $1/h$ with $h$ increase for the case of realistic $R_H = 10^{-8}$ Km$^2$/W; while the thick plateau or maximum appears for the ideal thermal contact case $R_H = 0$. The complex or oscillatory behavior of the phase $\mathrm{Arg}\left(\langle\delta T_G(z)\rangle\right)$ also occurs for the case $R_H = 0$ . The difference between the case $R_H = 0$ and $R_H \neq 0$ becomes smaller with frequency increase and disappears with at high frequencies $\omega\tau \geq 10$ (the case of high-frequency de-modulation). For the very high de-modulating frequencies $\langle\delta T_G(z)\rangle \sim 1/h$ independently on $R_H$ value.

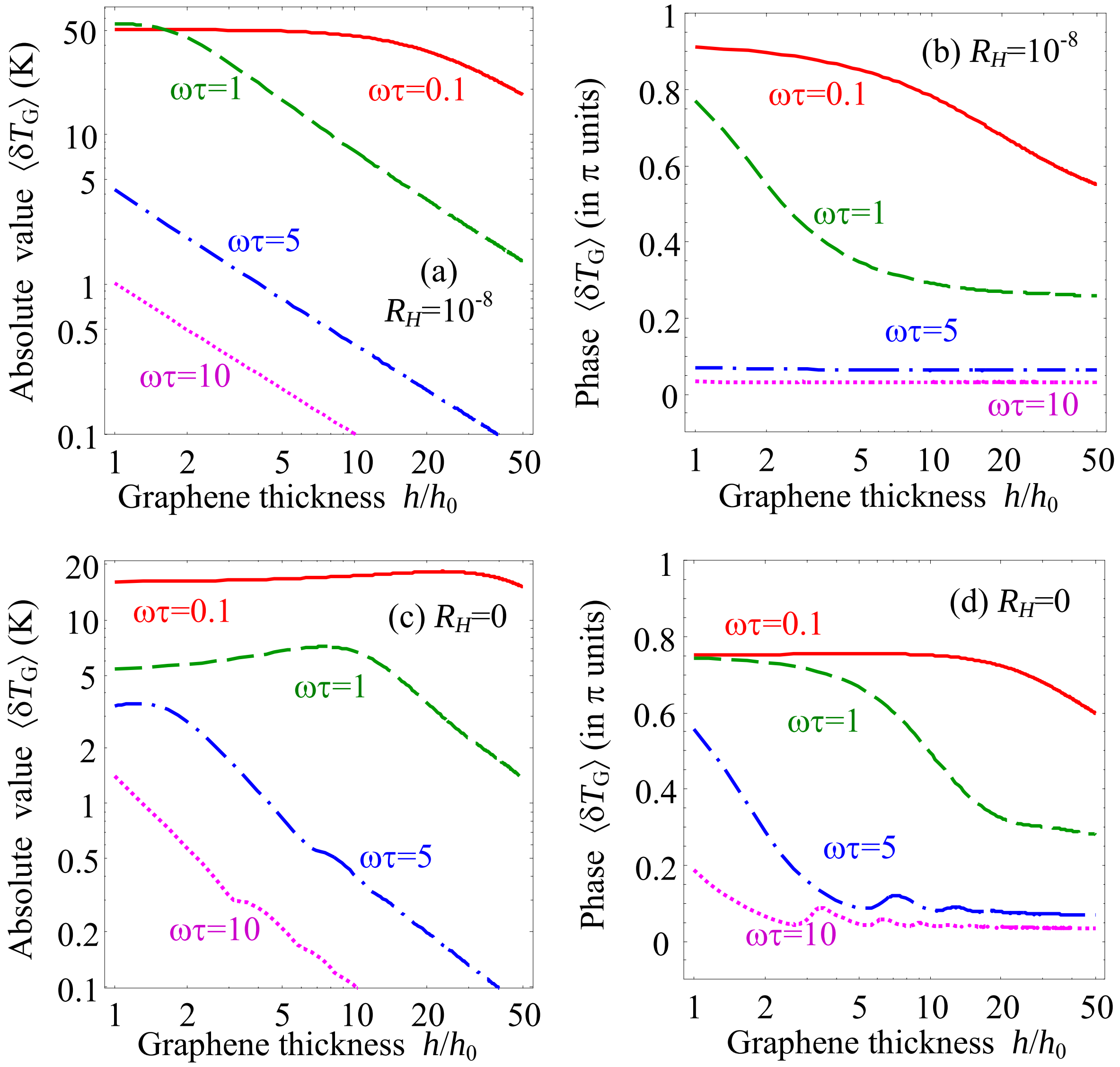

**Figure 3. Finite-size effect of the high-frequency temperature modulation.** Dependence of the averaged high-frequency temperature modulation absolute value $\left|\langle\delta T_G(z)\rangle\right|$ **(a,c)** and phase $\mathrm{Arg}(\langle\delta T_G(z)\rangle)$ **(b,d)** on graphene thickness $h/h_0$ calculated for dimensionless frequency $\omega\tau$ from 0.1 to 10 (as listed near the curves), interfacial resistance $R_H = 10^{-8}$ Km$^2$/W **(a, b)** and $R_H = 0$ **(c, d)**, electric field amplitude $E_0 = 5\times10^4$ V/m. Other parameters are the same as in the Figure 2.

The most interesting results presented in the section are (1) the strong impact of the Kapitza resistance on the dynamic temperature field in graphene and (2) the pronounced finite-size effect of the temperature modulation. The physical explanation of the result (1) is similar to that given for the static heating: Kapitza resistance try to block the extra heat inside graphene layers, but the blocking appeared effective only at frequencies $\omega\tau \leq 1$, lower than the high-frequency demodulation limit. At higher frequencies $\omega\tau \geq 5$ the ultra-thin heat gap leaks the heat flow, as electric capacitor leaks the high-frequency electric current. Unfortunately the explicit explanation of the finite-size effect (2) requires a microscopic treatment beyond the continuous approximation used here. The contradiction between the macroscopic description of the microscopic effect makes the size dependence complex including plateau or diffuse maxima, which width is frequency-dependent, followed by the scaling $\langle\delta T_G(z)\rangle \sim 1/h$ with thickness increase (see **Fig.3**).

## V. PYROELECTRIC MECHANISM OF THE CARRIER DENSITY MODULATION IN GRAPHENE

In the case, when the heating of graphene-on-ferroelectric by a high-frequency electric current is essential (e.g. of (1 − 10) K order or higher), a pronounced temperature gradient appears in ferroelectric substrate. Almost inertialess pyroelectric effect immediately transforms the gradient into the spontaneous polarization gradient [41] up to the frequencies $\omega\tau \leq 100$, since the pyro-reaction characteristic time is the Landau-Khalatnikov $\tau_{LK} \leq 10^{-13} s$, that is several order of magnitude smaller than the phonon relaxation time $\tau \approx 10^{-11} s$ . Assuming that the polarization direction is mainly perpendicular to the graphene-ferroelectric interface, a depolarizing electric field occurs and can penetrate in the multi-layer graphene. The field can be regarded as so-called “pyroelectric” field. In the assumption of good electric contact at the graphene-ferroelectric interface the graphene free charges immediately screen the pyroelectric field and thus their density oscillates at high-frequency in the vicinity of the interface.

It is worth to underline that the static heating of the single-domain ferroelectric substrate, given by Eqs.(10b) and (10d), can dominate in the heating effect for considered parameters. However the static polarization gradient induced by the static temperature gradient is typically almost screened by the sluggish “electrostatic” charge in ferroelectric (corresponding relaxation time is $\tau_{Mx} \geq 10^{-6} s$ ). Thus the static effect cannot create any noticeable free carrier density modulation in graphene. Only high-frequency modulation cannot be screened by the sluggish charge and can indeed modulate the electron density in graphene. Thus below we will consider only the high-frequency temperature modulation. The pyroelectric mechanism of the carrier density high-frequency modulation in graphene is illustrated in the **Figs. 4a,b**. Note, that the domain structure appearance in a ferroelectric substrate can lead to the in-plane carrier density modulation as illustrated in the **Fig. 4c.** The problem extension on the multi-domain case, which is important for large scale graphene based devices with a size greater than a single domain, imperfect electric contact or the electrostatic doping of graphene by random impurities as well as lateral heat dissipation is in progress now.

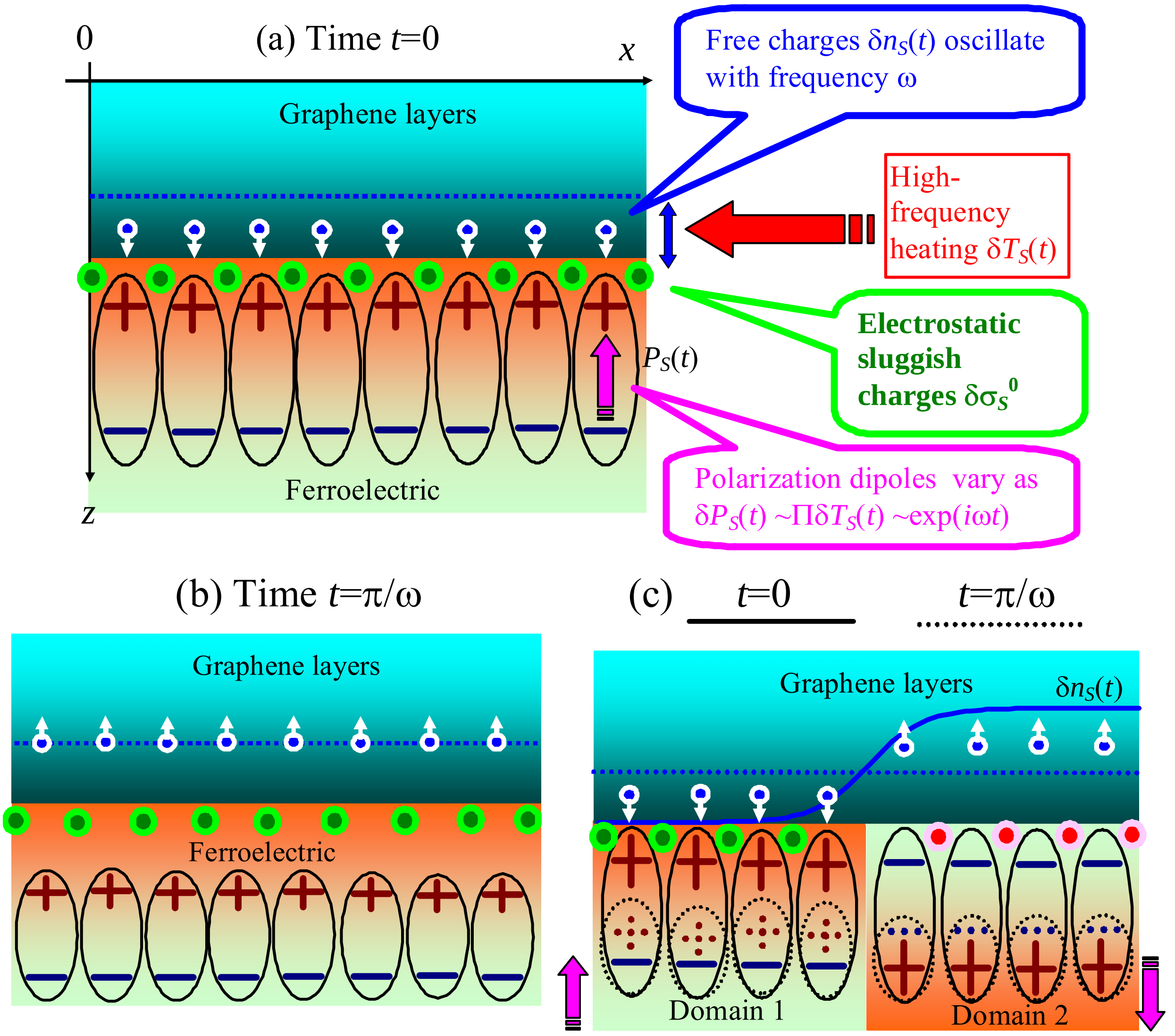

**Figure 4. Pyroelectric mechanism of the carrier density high-frequency modulation in graphene.** Free charge located at graphene-ferroelectric interface follows the high-frequency temperature variation due to the pyroelectric effect in the single-domain ferroelectric for the moment of time $t=0$ **(a)** and $t=\pi/\omega$ **(b)**. Free charge density variation $\delta n_S(t)\sim\langle\delta T_S(t)\rangle$. Spontaneous polarization is $P_S(t)$. Electrostatic charges are sluggish and cannot screen the high-frequency pyroelectric field. **(c)** Domain structure appearance in a ferroelectric substrate leads to the in-plane carrier density modulation.

Let us consider the case of the high concentration of carriers in the gated graphene, when they screen immediately and completely the out-of-plane ferroelectric spontaneous polarization. The spontaneous polarization $P_S(z,t)$ depends on the substrate temperature $T=T_0+T_S(z,t)$ according to Landau theory for ferroelectrics with the second type phase transition as $P_S(z,t)=\sqrt{\frac{\alpha_T}{\beta}\left(T_C-T(z,t)\right)}$.

Depolarization field $E_3(z,T)$ caused by the temperature gradient in an ideally electroded single-domain ferroelectric film with out-of-plane spontaneous polarization is [42]:

$$E_3(z,T)=\frac{\langle P_S(T(z))\rangle - P_S(T(z))}{\varepsilon_0\varepsilon_b}. \tag{11}$$

Here the average spontaneous polarization is $\langle P_S(T(z))\rangle=\frac{1}{L}\int_h^{L+h}P_S(T(z))dz$, $\varepsilon_0$ is a universal dielectric constant, $\varepsilon_b$ is a background permittivity [43]. The density of the surface charge located at the graphene-ferroelectric interface in order to screen the depolarization field $E_3(z,T)$ is [42]:

$$\sigma_S(t)=-\langle P_S(T(z))\rangle \approx -\langle P_S(T_0)\rangle + \Pi(T_0)\langle \delta T_S(z,t)\rangle, \tag{12}$$

where $\Pi(T)\equiv -\frac{\partial P_S(T)}{\partial T}$ is the pyroelectric constant of a given ferroelectric. Also we assume that $T_0 >> |\delta T_S(t)|$. Assuming that the modulation $\sigma_S(t)$ contributes into the density of surface states in linear mode, one can estimate the variation $\delta\sigma_S$ using expression (5b) for $T_S(z,t)$ as:

$$\delta\sigma_S=\delta\sigma_0^S+\delta\sigma_\omega^S(t), \tag{13a}$$

$$\delta\sigma_0^S=\Pi(T_0)\lambda\frac{e\mu n_{2D}E_0^2L}{4D_Gc_V} \tag{13b}$$

$$\delta\sigma_\omega^S(t)=\Pi(T_0)\lambda\frac{e\mu n_{2D}E_0^2\tau}{4c_Vh}\left(\frac{\theta_\omega k_G\left(1-e^{-k_SL}\right)}{\left(4\omega_0^2\tau^2-2i\omega_0\tau\right)k_S^2L}\sinh(k_Gh)e^{2i\omega_0t}+c.c.\right) \tag{13c}$$

As it was mentioned above, the stationary part of the variation, $\delta\sigma_0^S$, that may dominate in $\delta\sigma_S(t)$ effect for considered parameters, is screened by the sluggish free charge in ferroelectric and thus cannot create the electron density modulation in graphene. Thus below we will analyze the high-frequency modulation $\delta\sigma_\omega^S(t)$ only.

Equations (13) allow evaluating the impact of the pyroelectric effect. The value of the primarily pyroelectric constant $\Pi$ is 306 μC/m$^2$ K for Pb(Zr,Ti)O$_3$ at room temperature [44]. Therefore $\delta\sigma_\omega^S(t)=\Pi\langle\delta T_S(z,t)\rangle$ can be ~0.0003 C/m$^2$ for the frequencies that correspond temperature variation $\langle\delta T_S(z,t)\rangle$~1 K. The value of bulk spontaneous polarization is $P_S(T_0)$~0.5 C/m$^2$, i.e. the inequality $|P_S(T_0)| >> |\delta\sigma_S(t)|$ is valid with high accuracy and the variation of the ferroelectric polarization can be neglected. However, the corresponding variation of the carrier concentration in graphene, $\delta n_S(t)=\delta\sigma_\omega^S(t)/e\sim 10^{11}$ cm$^{-2}$, can be important for the low carrier concentrations in the gated graphene-on-ferroelectric. Note that the effect depends on the average temperature variation in ferroelectric substrate.

**Figure 5** presents the dependence of $\delta n_S(t)=\delta\sigma_\omega^S(t)/e$ on $N$-layer graphene thickness $N=h/h_0$ calculated for pyroelectric constant $\Pi$ = 306 μC/m$^2$ K, dimensionless frequency range ωτ=0.1 – 10, nonzero interfacial resistance $R_H$ and ideal thermal contact at graphene-ferroelectric interface. One can see that the case $R_H=0$ favors $\delta n_S(t)$ increase at frequencies $\omega\tau\geq 5$ due to the stronger temperature gradient in substrate in comparison with the realistic case $R_H=10^{-8}$ Km$^2$/W (compare **Figs. 5a,b** with **5c,d**). At moderate frequencies $\omega\tau\leq 1$ the interfacial heat resistance impact becomes small. The absolute value of $\delta n_S(t)$ is inversely proportional to the thickness $h$ for the case $N$>>10 at fixed gate voltage, in consistence with our model for the heat sources, where, the bulk 3D concentration should be substituted into conductivity σ. However $\delta n_S(t)$ behavior at $N$<10 can be non-monotonic and becomes oscillatory with frequency increase. Charge variation decreases with the increase of frequency and graphene thickness. At moderate frequencies 0.1 ≤ ωτ ≤ 1 the variation $|\delta n_S|$ is indeed more or about $10^{11}$ cm$^{-2}$ at $N$<20. At high frequencies ωτ ≥ 10 the variation $|\delta n_S|$ is less than $10^{10}$ cm$^{-2}$. Physically this correspond the case of high frequency demodulation due to the slow phonon relaxation in comparison with ac field variation.

**Figure 6** presents the frequency spectrum of carrier density, $\delta n_S(t)=\delta\sigma_\omega^S(t)/e$, calculated for several thicknesses $N=h/h_0$ of graphene, nonzero interfacial resistance $R_H$ and ideal thermal contact at graphene-ferroelectric interface. Concentration variation almost monotonically decreases with the ac-field frequency increase (or thickness increase) for nonzero resistance $R_H=10^{-8}$ Km$^2$/W (see **Figs. 6a,b**). Pronounced oscillations appear with the frequency increase on the background of the variation decrease for the ideal contact $R_H=0$ and $N<20$ (see **Figs. 6c,d**). The oscillation amplitude and period strongly depend on the number $N$ of graphene layers. The amplitude rapidly decreases and the period increases with $N$ increase. Corresponding analytical expression Eqs.(13c) contains the function $f(h,\omega,R_H)=\dfrac{\lambda\theta_\omega k_G\sinh(k_G h)}{k_S^2 Lh(4\omega_0^2\tau^2-2i\omega_0\tau)}$ that complexly depend on the graphene thickness $h$ and frequency ω. Dependence on the boundary conditions, e.g. on the Kapitza resistance $R_H$ and the heat conductivities $\lambda_{G,S}$, is included in the function $\theta_\omega(h,R_H)$. For a very high $R_H$ the asymptotic estimation $|f(h,\omega,R_H)|\propto\dfrac{1}{4\omega_0^2\tau^2(4\omega_0^2\tau^2+1)R_H\lambda_S hL}$ is valid, and so the absolute value monotonically decreases with frequency ω increase and/or thickness $h$ increase (no oscillations occurs). For the ideal heat contact ( $R_H=0$ ) and small ratio $\lambda=\lambda_G/\lambda_S<<1$ the function $\left|f(h,\omega,R_H)\right| \sim \frac{\lambda}{\omega_0^2\tau^2\left(4\omega_0^2\tau^2+1\right)Lh}\left|\frac{k_G}{\coth(k_G h)}\right|$ contain the oscillating factor $\left|\frac{k_G}{\coth(k_G h)}\right|$, since $k_G \sim \sqrt{2i\omega_0\tau - 4\omega_0^2\tau^2}$ is complex. The oscillatory behavior is absent for graphite ( $N \geq 50$ ) as anticipated ( $\coth(k_G h) \to 1$ ). So, the physical origin of the eventual carrier density oscillations is the mixed ballistic-Fourier mechanism of heat dissipation in multi-layer graphene exposed by the good heat contact of the graphene-ferroelectric interface. Note, that the oscillatory resonant-like behavior of the carrier density spectrum can be very interesting for excitation and enhancement of the eigen modes along the interface. However, the realization of this behavior needs interfaces with low Kapitza resistances.

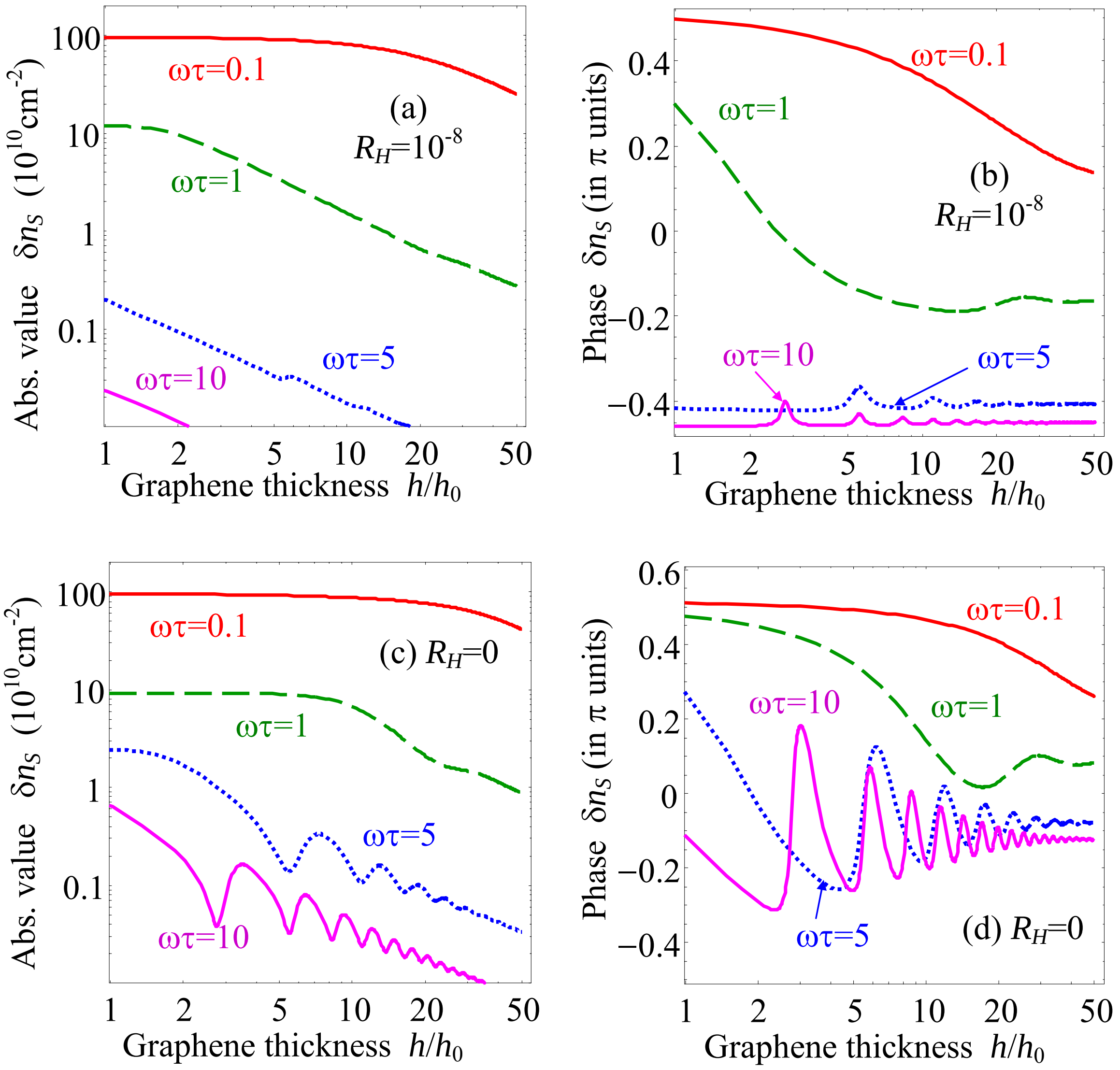

**Figure 5. Finite-size effect of carrier density modulation in multi-layer graphene.** Dependence of the concentration variation absolute value $|\delta n_S|$ **(a,c)** and phase $\text{Arg}(\delta n_S)$ **(b,d)** on graphene thickness $N = h/h_0$ calculated for dimensionless frequency $\omega\tau$ from 0.1 to 10 (as listed near the curves),

interfacial resistance $R_H$ = $10^{-8}$ Km$^2$/W **(a, b)** and $R_H$ = 0 **(c, d)**, pyroelectric constant $\Pi = 306$ μC/m$^2$ K and electric field amplitude $E_0 = 5\times10^4$ V/m. Other parameters are the same as in the Figure 2.

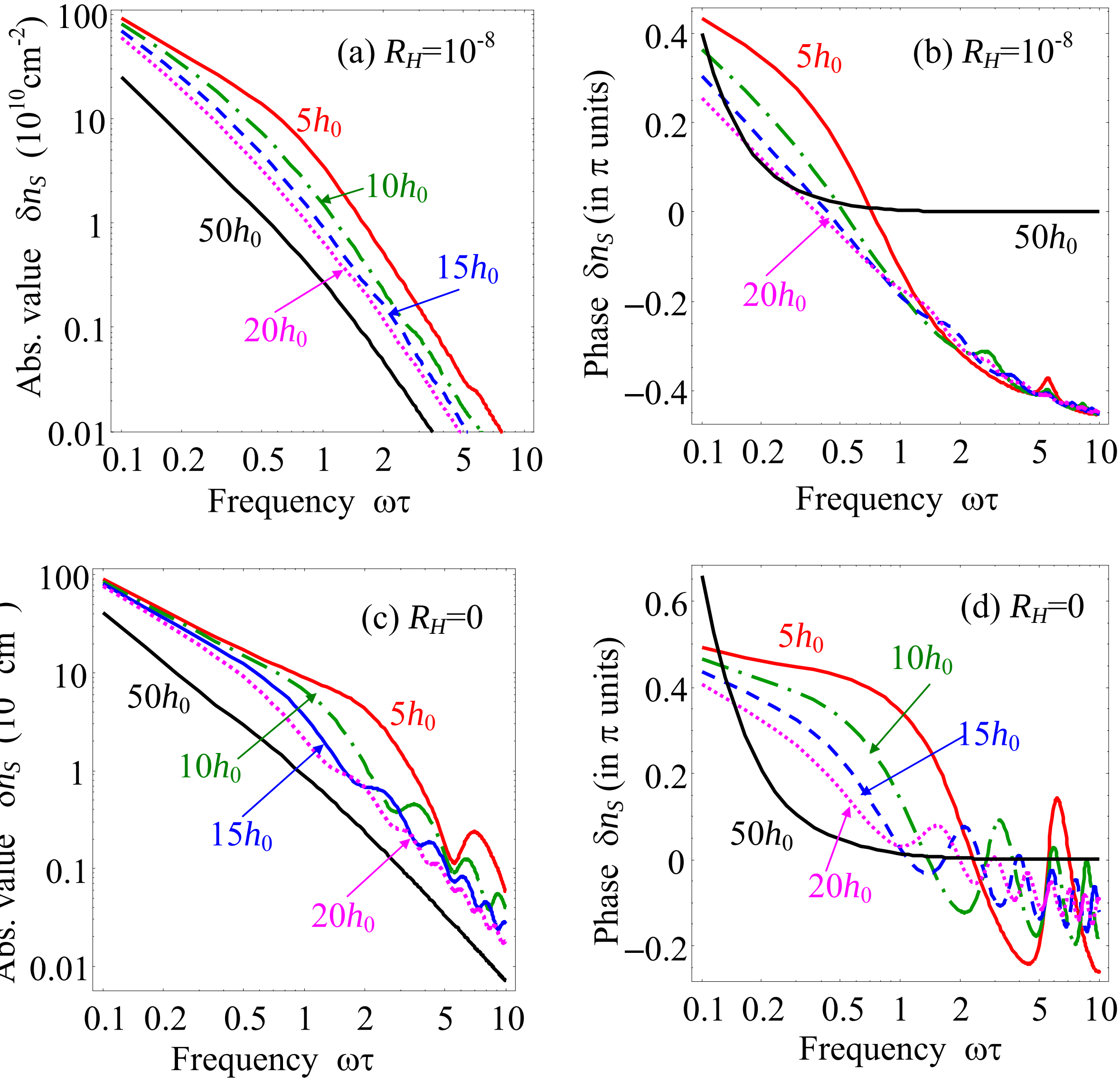

**Figure 6. Frequency spectrum of carrier density variation in multi-layer graphene.** Dependence of the concentration variation absolute value $|\delta n_S|$ **(a,c)** and phase $\mathrm{Arg}(\delta n_S)$ **(b,d)** on dimensionless frequency ωτ calculated for graphene thickness $N = h/h_0$ from 5 to 50 (as listed near the curves), interfacial resistance $R_H$ = $10^{-8}$ Km$^2$/W **(a, b)** and $R_H$ = 0 **(c, d)**, pyroelectric constant $\Pi = 306$ μC/m$^2$ K and electric field amplitude $E_0 = 5\times10^4$ V/m. Other parameters are the same as in the Figure 2.

Contour map of the concentration variation absolute value $|\delta n_S|$ in coordinates "graphene thickness $h$ - frequency $\omega\tau$" is shown in **Figure 7.** The 2D-density of carriers variation, $\delta n_S(t) \sim 10^{12}$ cm$^{-2}$, is essential in comparison with the 2D concentration of carriers in graphene layer at the small gate voltages, but the charge variation $\delta\sigma_S(t) \sim 3\times10^{-3}$ C/m$^2$ is negligibly small in comparison with spontaneous polarization of $Pb(Zr,Ti)O_3$ substrate, which permits to neglect non-linearity of the heating.

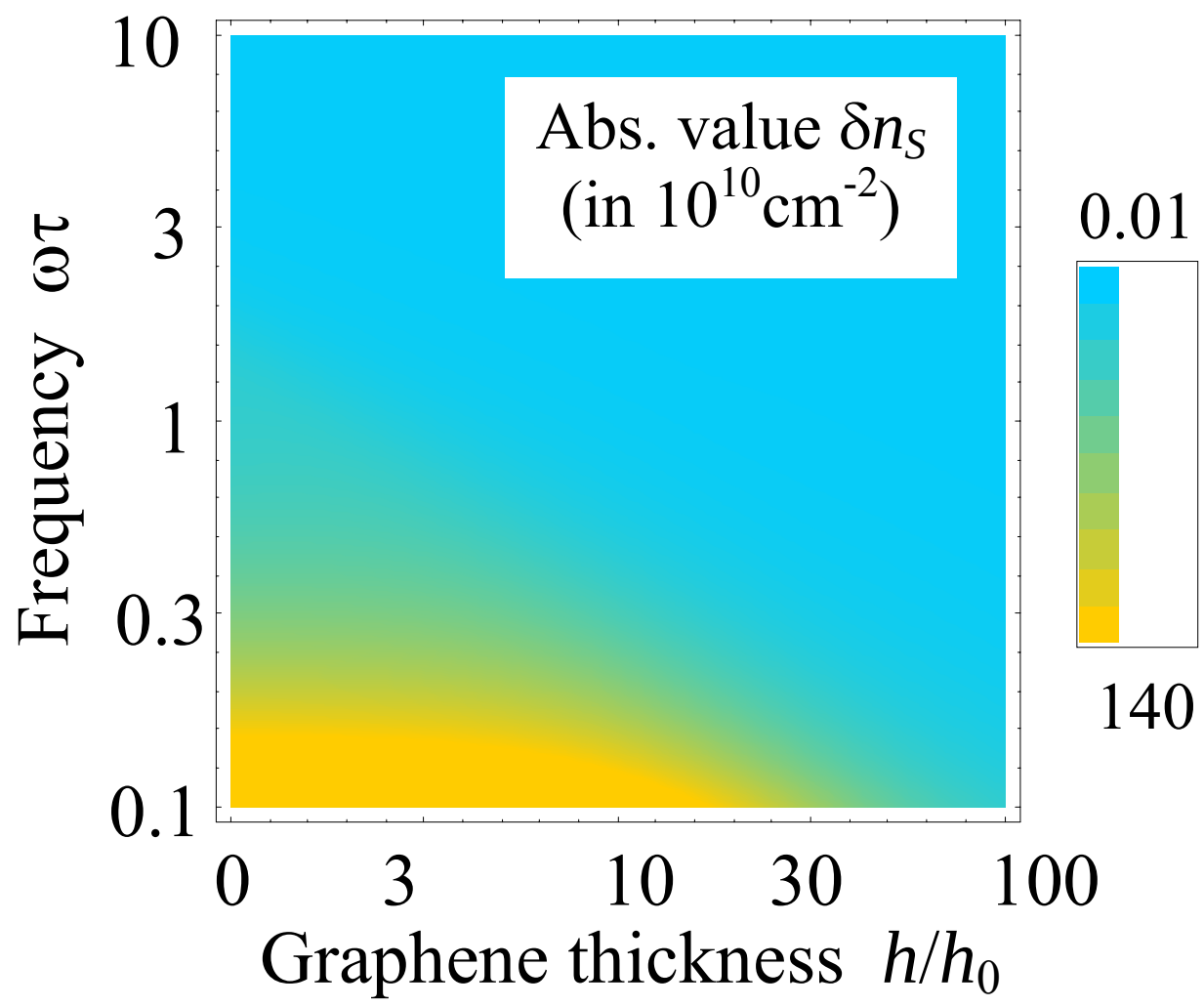

**Figure 7.** Contour map of the concentration variation absolute value $|\delta n_S|$ in coordinates thickness $h/h_0$ - frequency $\omega\tau$. Interfacial resistance $R_H = 10^{-8}$ Km$^2$/W; other parameters are the same as in the Figure 5.

One can conclude from the section that the carrier concentration modulation in graphene $\delta n_S(t) \sim \langle \delta T(z,t) \rangle$ caused by the pyroelectric effect can modify essentially within one period, demonstrates a pronounced finite-size effect, non-monotonic or oscillatory frequency spectrum and complex dependence on the interfacial resistance.

## VI. DISCUSSION

Heat dissipation in multi-layer graphene-on-ferroelectric has been studied within a continuous approximation for the heat transfer across the graphene layers. We demonstrate that the Joule effect, caused by a high-frequency ac electric current in graphene, creates a pronounced temperature gradient in a ferroelectric substrate. The static heating of the graphene layers strongly depends on the value of interfacial Kapitza resistance. The interfacial resistance blocks all "extra" heat in graphene acting as effective heat gap. The optimal amount of heat is determined by the heat flux continuity at the graphene-ferroelectric interface; it causes the linear temperature gradient in the ferroelectric film. The static heating of ferroelectric does not depend on the Kapitza resistance as well as on the graphene thickness.

The most interesting results we obtained are the strong impact of the Kapitza resistance on the high-frequency temperature field in graphene and the pronounced finite-size effect of the temperature modulation. In particular Kapitza resistance, acting as the ultra-thin heat gap, try to block the extra heat inside graphene layers, but the blocking appeared to be effective only at frequencies less than 100 GHz. At THz-frequencies the gap leaks the heat flow, as electric capacitor leaks the high-frequency electric current (the effect of high frequency demodulation). The finite-size effect is complex, corresponding size dependence includes plateau or diffuse maxima, with a frequency-dependent width, followed by the scaling 1/h with multi-layer graphene thickness *h* increase.

Pyroelectric effect impact was taken into consideration within Landau theory of ferroelectrics. The pyroelectric effect transforms the temperature gradient into the spontaneous polarization gradient. In assumption that the spontaneous polarization is perpendicular to the graphene-ferroelectric interface, the high-frequency depolarizing electric field occurs and penetrates in the multi-layer graphene. Free charges in graphene immediately screen the electric field and thus their density oscillates at high-frequency. In particular the electron concentration variation is proportional to the charge variation induced by the temperature gradient in ferroelectric substrate via the primary pyroelectric effect. The carrier density modulation in graphene, caused by the pyroelectric mechanism, demonstrates the strong finite-size effect, non-monotonic or oscillatory frequency spectrum and complex dependence on the interfacial resistance. The physical origin of the carrier density oscillations is caused by the mixed ballistic-Fourier mechanism of heat dissipation in a multi-layer graphene exposed by the good heat contact at the graphene-ferroelectric interface. Note, that the oscillatory resonant-like behavior of the carrier density spectrum can be very promising for excitation and enhancement of the eigen modes along the graphene-ferroelectric interface.

Calculations proved that the pyroelectric effect can modify essentially the free carrier density at the graphene-ferroelectric interface and consequently its conductivity. So, pyroelectric mechanism can be critical for understanding of the complex physical processes taking place across graphene-ferroelectric interface at terahertz frequencies.

Pyroelectric mechanism and finite-size effects should be taken into consideration for the non-volatile memory devices and modulators, based on multi-layer graphene, placed over the ferroelectric substrate, operating at high frequencies [3, 7-12] of 100 GHz order and lower, because it can possibly limit the switching rate of such devices. Our examination demonstrates that the suitable range of comparatively moderate concentrations should be chosen in order to make the devices more fast, although this would reduce somewhat the graphene channel conductivity.

Note, that continuous model of heat transport, used in our study, impose the restrictions for the range of multi-layer graphene with a few layers only. Obtained results correspond to the case of $N$-layer graphene with thickness of the diffusion length order or greater. This makes them suitable for the range of 5-7 or more graphene layers. However, this very case can be important from the point of fundamental studies and practical applications. The case of small graphene thickness needs microscopic *ab initio* theory, which is beyond the limits of this our work.

**Acknowledgements**

Authors gratefully acknowledge critical remarks and multiple discussions with E.A. Eliseev. This work was partially supported by the State Fund of Fundamental Research of Ukraine.

## SUPPLEMENTAL MATERIALS

## APPENDIX A

We shall search for the periodical response at ω frequency and the dependence of sources density on *z* (the last can be verified for the case of $N$ >>1, when multi-layer graphene transforms in fact into the ultra-thin graphite). Joule heating of graphene occurs due to electric current, in turn caused by the field $E(t) = E_0 \sin(\omega_0 t)$, that creates heat sources with density

$$q(t) = \sigma\left|E^2(t)\right| = \frac{\sigma E_0^2}{2}\left(1 - \cos(2\omega_0 t)\right) \tag{A.1a}$$

σ is bulk conductivity of multi-layer graphene. Below symbol "~" over a letter stands for its frequency Fourier image:

$$\tilde{q}(\omega) = \frac{\sigma E_0^2}{2}\left(\delta(\omega) - \frac{\delta(\omega - 2\omega_0)}{2} - \frac{\delta(\omega + 2\omega_0)}{2}\right), \tag{A.1b}$$

where $\delta(\omega)$ is Dirac-delta function.

In Fourier domain we are looking for the solution of the three-layered vacuum-graphene-substrate thermal problem:

$$\omega^2\tilde{T} - \frac{i\omega}{\tau}\tilde{T} + \frac{D_G}{\tau}\frac{d^2\tilde{T}}{d^2 z} = -\frac{\sigma E_0^2}{2\tau c_V}\left(\delta(\omega) - \frac{\delta(\omega - 2\omega_0)}{2} - \frac{\delta(\omega + 2\omega_0)}{2}\right), \quad 0 \le z \le h, \tag{A.2a}$$

$$-i\omega\tilde{T} + D_S\frac{d^2\tilde{T}}{d^2 z} = 0, \qquad h < z < L. \tag{A.2b}$$

With the boundary conditions of the thermal flux absence in vacuum, $\frac{d}{dz}\tilde{T}(0) = 0$, thermal flux continuity the graphene-substrate interface, $\lambda_G \frac{d}{dz}\tilde{T}(h-0) = \lambda_S \frac{d}{dz}\tilde{T}(h+0)$, temperature jump related with Kapitza resistance, $\tilde{T}(h-0) - \tilde{T}(h+0) = R_H Q \equiv -R_H \lambda_G \frac{d}{dz}\tilde{T}(h)$ , and unperturbed constant temperature field far from the interface in substrate. Using the expression for graphene resistance, $R_G = \left(\tilde{T}_G(0) - \tilde{T}_G(h)\right)/\langle Q\rangle \equiv h/\lambda_G$ , for estimations it makes sense to rewrite the condition $\tilde{T}(h-0) - \tilde{T}(h+0) = -R_H \lambda_G \frac{d}{dz}\tilde{T}(h)$ as $\tilde{T}_G(h) + h\frac{R_H}{R_G}\frac{d}{dz}\tilde{T}_G(h) = \tilde{T}_S(h)$. The heat penetration thickness $L$ is regarded much greater than the temperature wavelength at high frequency $2\omega_0$, but much smaller than the thickness for which a linear temperature gradient in substrate cannot realized in the static limit $\omega_0 = 0$.

General solution of Eq.(A.2) is

$$T(z) = T_0(z) + T_\omega(z)e^{2i\omega_0 t} + c.c. \qquad \text{(A.3)}$$

where

$$T_0(z) = \begin{cases} -\dfrac{\sigma E_0^2}{4D_G c_V} z^2 + A_1 + A_2 z, & 0 \le z \le h, \\ B_1 + B_2 z, & h < z < L \end{cases} \qquad \text{(A.4a)}$$

Constants $A_{1,2}$ and $B_{1,2}$ can be found from the aforementioned boundary conditions:

$$\begin{aligned}
&\frac{d}{dz}\tilde{T}(0) = 0 \quad \Rightarrow \quad A_2 = 0, \\
&\lambda_G \frac{d}{dz}\tilde{T}(h) = \lambda_S \frac{d}{dz}\tilde{T}(h) \quad \Rightarrow \quad \frac{\lambda_G}{\lambda_S}(2S_0 h + A_2) = B_2 \quad \to \quad B_2 = +\frac{\lambda_G}{\lambda_S} 2S_0 h, \\
&\tilde{T}(h-0) - \tilde{T}(h+0) = -R_H \lambda_G \frac{d}{dz}\tilde{T}(h) \quad \Rightarrow \quad S_0 h^2 + A_1 + A_2 h - B_1 - B_2 h = -R_H \lambda_G (2S_0 h + A_2) \\
&\to \quad A_1 = -S_0 h^2 + B_1 + B_2 h - R_H \lambda_G 2S_0 h \\
&\tilde{T}(L) = T_0 \quad \Rightarrow \quad (L)B_2 + B_1 = T_0
\end{aligned}$$

(A.4b)

Thus the stationary part of solution is:

$$T_0(z) = \begin{cases} T_0 - \dfrac{\sigma E_0^2}{4D_G c_V} z^2 + \dfrac{\sigma E_0^2}{4D_G c_V} h\left( h + 2\dfrac{\lambda_G}{\lambda_S} L + 2R_H \lambda_G \right), & 0 \le z \le h, \\ T_0 + \dfrac{\lambda_G}{\lambda_S} \dfrac{\sigma E_0^2}{2D_G c_V} h(L + h - z), & h < z \le L + h \end{cases} \qquad \text{(A.5)}$$

It is seen from Eq.(A.5) that the solution increases with $L$ increase and becomes unphysical at $L \to \infty$. Fourier spectrum of the high-frequency part of the solution is:

$$\tilde{T}_\omega(z) = \begin{cases} \dfrac{\sigma E_0^2}{4c_V} \dfrac{\delta(\omega - 2\omega_0)}{\omega^2 \tau - i\omega} + C_1 \exp(k_G z) + C_2 \exp(-k_G z), & 0 \le z \le h, \\ C_0 \exp(-k_S z), & h < z \end{cases} \qquad \text{(A.6)}$$

Where $k_G(\omega) = \sqrt{\dfrac{i\omega - \omega^2 \tau}{D_G}} \equiv \sqrt{\dfrac{i\omega\tau - \omega^2 \tau^2}{D_G \tau}}$ and $k_S(\omega) = (1+i)\sqrt{\dfrac{\omega}{2D_S}}$ .Constant $C_{0,1,2}$ can be found from the aforementioned boundary conditions:

$$\begin{aligned}
&C_1 - C_2 = 0, \\
&S_\omega + C_1 \exp(k_G h) + C_2 \exp(-k_G h) - C_0 \exp(-k_S h) = R_H \lambda_G (C_1 \exp(k_G h) - C_2 \exp(-k_G h)), \qquad \text{(A.7)} \\
&\frac{\lambda_G}{\lambda_S} k_G (C_1 \exp(k_G h) - C_2 \exp(-k_G h)) = -k_S C_0 \exp(-k_S h).
\end{aligned}$$

Where $S_\omega = \dfrac{\sigma E_0^2}{4c_V} \dfrac{\delta(\omega - 2\omega_0)}{\omega^2 \tau - i\omega}$. The solution is

$$\widetilde{T}_{\omega}(z)=\begin{cases} S_{\omega}-\dfrac{S_{\omega}k_S\cosh(k_G z)}{k_S\cosh(k_G h)+\lambda k_G(1+R_H k_S\lambda_S)\sinh(k_G h)}, & 0\le z\le h,\\ \dfrac{S_{\omega}k_G\lambda\sinh(k_G h)\exp(k_S(h-z))}{k_S\cosh(k_G h)+\lambda k_G(1+R_H k_S\lambda_S)\sinh(k_G h)}, & h<z \end{cases} \qquad \text{(A.8)}$$

Where $\lambda=\lambda_G/\lambda_S$ . The final answer was derived in accordance with Eq.(A.3), (A.5) and (A.8). Corresponding temperature field in graphene ( $0\le z\le h$ ) is:

$$\begin{aligned} T_G(z,t)=T_0&+\frac{\sigma E_0^2 h^2}{4D_G c_V}\left(1-\frac{z^2}{h^2}+2\lambda\frac{L}{h}+2R_H\frac{\lambda_G}{h}\right)\\ &+\frac{\left(\sigma E_0^2\tau/4c_V\right)}{\left(4\omega_0^2\tau^2-2i\omega_0\tau\right)}\left(\exp(2i\omega_0 t)-\frac{k_S\cosh(k_G z)\exp(2i\omega_0 t)}{k_S\cosh(k_G h)+\lambda k_G(1+R_H k_S\lambda_S)\sinh(k_G h)}\right)+c.c. \end{aligned} \qquad \text{(A.9a)}$$

The temperature field in a ferroelectric film ( $h<z<L+h$ ) is:

$$T_S(z,t)=T_0+\frac{\sigma E_0^2 h^2}{2D_G c_V}\lambda\frac{L+h-z}{h}+\frac{\left(\sigma E_0^2\tau/4c_V\right)}{\left(4\omega_0^2\tau^2-2i\omega_0\tau\right)}\frac{k_G\lambda\sinh(k_G h)\exp(k_S(h-z))\exp(2i\omega_0 t)}{k_S\cosh(k_G h)+\lambda k_G(1+R_H k_S\lambda_S)\sinh(k_G h)}+c.c. \qquad \text{(A.9b)}$$

Parameters $k_G=\sqrt{\dfrac{2i\omega_0-4\omega_0^2\tau}{D_G}}\equiv\sqrt{\dfrac{2i\omega_0\tau-4\omega_0^2\tau^2}{D_G\tau}}$ and $k_S=\dfrac{(1+i)}{2}\sqrt{\dfrac{2\omega_0}{D_S}}$ . It is worth to underline that the stationary temperature distribution in the ferroelectric is independent on the Kapitza resistance.

Equation (A.9b) gives expression for $\langle T_S(z,t)\rangle$ and the explicit form:

$$\langle T_S(z,t)\rangle\approx\frac{\sigma E_0^2 hL}{4D_G c_V}\lambda+\frac{\left(\sigma E_0^2\tau/8c_V\right)}{\left(2\omega_0^2\tau^2-i\omega_0\tau\right)}\frac{k_G\lambda\sinh(k_G h)(1-\exp(-k_S L))\exp(2i\omega_0 t)}{k_S L\left(k_S\cosh(k_G h)+\lambda k_G(1+R_H k_S\lambda_S)\sinh(k_G h)\right)}+c.c. \qquad \text{(A.10)}$$

$\delta T$ profile in multi-layer graphene of thickness $h=15h_0$ placed on ferroelectric substrate is shown in the **Figure S1.**

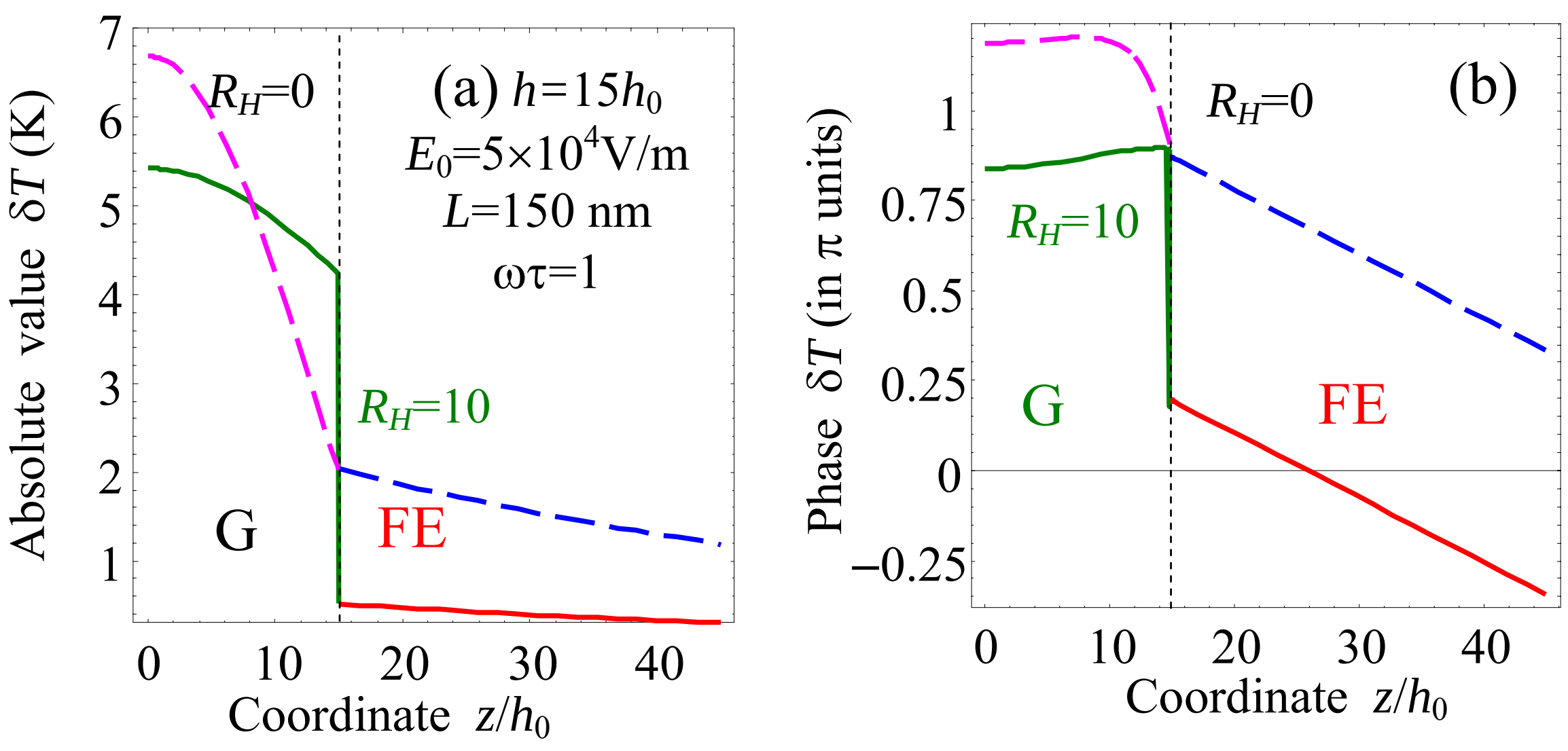

**Figure S1. Dynamic heating.** $\delta T$ profile in multi-layer graphene (G) of thickness $h = 15h_0$ (5.1 nm) and ferroelectric substrate (FE). Absolute value $|\delta T(z)|$ **(a)** and phase $\mathrm{Arg}(\delta T(z))$ **(b)** were calculated for dimensionless frequency $\omega\tau=1$, interfacial resistance $R_H = 10^{-8}$ Km²/W (solid curves) and $R_H = 0$ (dashed curves), electric field amplitude $E_0 = 5\times10^4$ V/m, parameters $D_G/D_S$ =0.1, $\lambda_G/\lambda_S$ = 0.1, $l_D \approx 3$ nm, $L = 150$ nm, $\lambda_G = 1$ W/(m·K) and $E_0 = 5\times10^4$ V/m. Other material parameters are listed in the **Table SI**.

**Table SI.** Material parameters used in calculations

| **Multi-layer graphene** | **Parameters used in our calculations** | **Typical range and/or references** |
|---|---|---|
| heat capacity $c_V$ | $1.534\times10^6$ J/m³K | [37] |
| Coefficient $K$ | 2 W/mK | 2 – 20 W/mK across the graphene layers [10] |
| relaxation time $\tau$ | $10^{-11}\,s$ | [38] |
| thermal diffusion $D_G$ | $0.13\cdot10^{-5}\,\mathrm{m}^2/s$ | $(0.13-1.3)\cdot10^{-5}\,\mathrm{m}^2/s$ calculated as $D_G = K/c_V$ |
| heat conductivity $\lambda_G$ | 1 W/(m·K) | ≥1 W/(m·K) [10, 32] |
| diffusion length $l_D$ | 3 nm | $(3-10)$ nm, calculated as $l_D = \sqrt{D_G\tau}$ |
| conductivity $\sigma(h)$ | $e\mu n_{2D}/h$, $n_{2D} \approx 10^{18}\,\mathrm{m}^{-2}$ | [3, 27] |
| mobility $\mu$ | $1.4\times10^1$ m²/Vs | [13] |
| thickness | $h_0 = 0.34$ nm, $h = Nh_0$, | $N = 5,\ldots 50$ |
| **Interface** | | |
| interfacial resistance | $R_H = 0,\ 10^{-9},\ 5\times10^{-9},\ 10^{-8}$ Km²/W | $R_H = (0.5-1)10^{-8}$ Km²/W [10] |
| **Ferroelectric** | | |
| Thickness $L$ | 100 nm | 30 – 300 nm [11-16] |
| thermal diffusion $D_S$ | $10D_G$ | $D_G << D_{FE}$ |
| heat conductivity $\lambda_S$ | $10\lambda_G$ | $\lambda_G << \lambda_{FE}$ |
| pyroelectric coefficient $\Pi$ | 306 μC/m² K (PZT) | [44, Table 1] |
| ***ac* field parameters** amplitude $E_0$ | (0.1 – 5)×10⁴ V/m | transport in graphene is linear up to 2×10⁵ V/m [40] |
| frequency range | $0.1 \le \omega\tau \le 10$ | $\omega_0$~100 GHz – 1 THz |